\def\mearth{{\rm\,M_\oplus}}

\documentclass[12pt]{article}

\usepackage{scicite}

\usepackage{graphicx}
\usepackage{deluxetable}

\usepackage{times}

\newenvironment{sciabstract}{%
\begin{quote} \bf}
{\end{quote}}

\newcounter{lastnote}

\title{Exotic Earths: Forming Habitable Worlds with Giant Planet Migration}

\author
{Sean N. Raymond,$^{1,4,5\ast}$ Avi M. Mandell,$^{2,3,4,5\ast}$ Steinn Sigurdsson$^{2,4}$\\
\\
\normalsize{$^{1}$Laboratory for Atmospheric and Space Physics, University of Colorado}\\
\normalsize{Boulder, CO 80309-0392}\\
\normalsize{$^{2}$Department of Astronomy and Astrophysics, Pennsylvania State University}\\
\normalsize{University Park, PA 16802}\\
\normalsize{$^{3}$NASA Goddard Space Flight Center}\\
\normalsize{Greenbelt, MD 20771}\\
\normalsize{$^{4}$Member of NASA Astrobiology Institute.}\\
\normalsize{$^{5}$S.R. and A.M. contributed equally to this research.}\\
\\
\normalsize{$^\ast$To whom correspondence should be addressed; E-mail:} \\
\normalsize{raymond@lasp.colorado.edu, mandell@astro.psu.edu.}  }

\date{}

\begin{document} 


\baselineskip24pt


\maketitle 


\begin{sciabstract}
Close-in giant planets (e.g. ``Hot Jupiters'') are thought to form far from
their host stars and migrate inward, through the terrestrial planet zone, via
torques with a massive gaseous disk. Here we simulate terrestrial planet
growth during and after giant planet migration.  Several-Earth mass planets
also form interior to the migrating Jovian planet, analogous to
recently-discovered ``Hot Earths''.  Very water-rich, Earth-mass planets form
from surviving material outside the giant planet's orbit, often in the
Habitable Zone and with low orbital eccentricities.  More than a third of the
known systems of giant planets may harbor Earth-like planets.
\end{sciabstract}



To date, giant planets have been detected around almost 200 main-sequence
stars \cite{but06,sch02}. An unexpected result is the abundance of planets
very close to their host stars -- about 40\% of the known extrasolar planets
are interior to Mercury's orbital distance of 0.4 Astronomical Units (AU; 1 AU
is the Sun-Earth distance), although observational biases favor the detection
of hot Jupiters \cite{tab02}.  The occurence of close-in giant planets is
surprising because models predict that giant planets form much more easily in
the cold outer regions of protoplanetary disks \cite{bod00,bos97}.  These
planetary systems have been attributed to inward migration of a giant planet
on 10$^5$ year timescales caused by an imbalance of torques generated by the
gaseous protoplanetary disk \cite{lin86,lin96,dang03,note1}.  In the process,
the giant planet moves through the terrestrial planet zone (located from a few
tenths of an AU to about 2-3 AU).  Radioactive dating of Solar System material
\cite{kl02} and observations of dust dispersal in disks around young stars
\cite{sil06} indicate that rapid precipitation and coagulation of solid
material in the inner regions of circumstellar disks is likely, leading to the
question of the fate of these protoplanets during and after giant planet
migration.  Previous studies on the possibility of Earth-like planets
co-existing with close-in giant planets are divided
\cite{arm03,man03,edg04,rql05,fog05}.

Here we simulate the growth and dynamical evolution of protoplanetary material
from small bodies to terrestrial planets during and after the migration of a
giant planet through the terrestrial zone (see supporting online material for
details). Simulations start from a circumstellar disk in the middle stages of
planet formation, extending from 0.25 to 10 AU. The disk contains 17 Earth
masses ($\mearth$) of rocky/icy material, evenly divided between 80 Moon- to
Mars-sized ``planetary embryos''\cite{wet89} and 1200 ``planetesimals'' with
properties modified so that each body behaves as collection of less massive
objects \cite{tho03}. The disk has a compositional gradient: the inner disk is
iron-rich and water-poor while the outer disk is water-rich and iron-poor (as
in \cite{rql04} but with 50\% water by mass beyond 5 AU.).  A Jupiter-mass
giant planet starts at 5 AU and is migrated in to 0.25 AU in 10$^5$ years
\cite{dang03}. The orbits of all bodies in each simulation are integrated for
200 million years with the hybrid symplectic integrator Mercury \cite{cha99},
modified to include two additional effects: 1) ``type 2'' giant planet
migration \cite{man03,lin86} and 2) aerodynamic gas drag \cite{tho03} from a
gaseous disk which dissipates on a 10$^7$ year timescale \cite{hai01}.

At early times (Fig. 1) the giant planet migrates inward through the disk,
causing nearby material to either be scattered outward onto high-eccentricity
orbits \cite{man03} or shepherded inward by the giant planet's moving
mean-motion resonances \cite{tka99}. The buildup of inner material induces
rapid growth of a four Earth mass planet just inside the 2:1 mean motion
resonance in 10$^5$ years (also shown by \cite{fog05}). Smaller bodies
(planetesimals) feel a stronger drag force and are shepherded by higher-order
resonances (in this case, the 8:1) and form a pileup of 0.2 $\mearth$ at 0.06
AU. At the end of the migration period, the remaining disk material is divided
between bodies captured in low eccentricity orbits in interior resonances with
the Jupiter-mass planet and higher eccentricity orbits beyond 0.5 AU. The
protoplanetary disk is now dynamically hot (i.e. orbital eccentricities and
inclinations are high), and accretion proceeds at a slower rate than would
occur in a non-stirred, dynamically cold disk. However, the gas continues to
damp eccentricities and inclinations, also causing the orbits of icy
planetesimals from the outer disk to decay inwards on million year timescales,
delivering a large amount of water to the growing terrestrial planets. After
the gas dissipates (at 10$^7$ years), the disk is stirred by interactions
between bodies and clearing continues through scattering. After 200 million
years the inner disk is comprised of the collection of planetesimals at 0.06
AU, a four Earth-mass planet at 0.12 AU, the Hot Jupiter at 0.21 AU, and a 3
Earth-mass planet at 0.91 AU.  Previous results have shown that these planets
are likely to be stable for billion year timescales \cite{rql05}.  Many
bodies remain in the outer disk, and accretion and ejection are ongoing due to
long orbital timescales and high inclinations.

Two of the four simulations from Fig. 2 shown contain a $>$0.3 $\mearth$
planet on a low-eccentricity orbit in the Habitable Zone, where the
temperature is adequate for water to exist as liquid on a planet's surface
\cite{kas93}.  We adopt 0.3 $\mearth$ as a lower limit for habitability,
including long-term climate stabilization via plate tectonics \cite{wil97}.
The surviving planets can be broken down into three categories: 1) Hot Earth
analogs interior to the giant planet, 2) ``normal'' terrestrial planets
between the giant planet and 2.5 AU, and 3) outer planets beyond 2.5 AU, whose
accretion has not completed by the end of the simulation.  Properties of
simulated planets are segregated (Table 1): Hot Earths have very low
eccentricities and inclinations and high masses because they accrete on the
migration timescale (10$^5$ years), so there is a large amount of damping
during their formation.  These planets are reminiscent of the recently
discovered, close-in 7.5 $\mearth$ planet around GJ 876 \cite{riv05}, whose
formation is also attributed to migrating resonances \cite{zho05}.  Farther
from the star, accretion timescales are longer and the final phases take place
after the dissipation of the gas disk (at 10$^7$ years), causing the outer
terrestrials to have large dynamical excitations and smaller masses, because
accretion has not completed by 200 million years; collisions of outer bodies
such as these may be responsible for dusty debris disks seen around
intermediate-age stars \cite{kim05}. In the ``normal'' terrestrial zone,
dynamical excitations and masses fall between the two extremes as planets form
in a few times 10$^7$ years, similar to the Earth's formation
timescale\cite{kl02}.  In addition, the average planet mass in the terrestrial
zone is comparable to the Earth's mass, and orbital eccentricities are
moderate (Table 1).

Both the Hot Earths and outer Earth-like planets have very high water contents
(up to $>$100 times that of Earth\cite{note2}) and low iron contents compared
with our own terrestrial planets (Table 1). There exist two sources for these
trends in composition: 1) strong radial mixing induced by the migrating giant
planet, and 2) an influx of icy planetesimals from beyond 5 AU from gas
drag-driven orbital decay that is unimpeded by the scattering that Jupiter
performs in our own system. The outer terrestrial planets acquire water from
both of these processes, but the close-in giant planet prevents in-spiraling
icy planetesimals from reaching the Hot Earths.  The accretion of outer,
water-rich material dilutes the high iron content of inner disk material, so
water-rich bodies naturally tend to be iron-poor in terms of mass fraction.
The high water contents of these planets suggest that their surfaces would be
most likely covered by global oceans several km deep.  Additionally, their low
iron contents may have consequences for the evolution of atmospheric
composition \cite{cla06}.

The spacing of planets (Fig. 2) is highly variable -- in some cases planets
form relatively close to the inner giant planet.  The ratio of orbital periods
of the innermost $>$0.3 $\mearth$ terrestrial planet to the close-in giant
ranges from 3.3 to 43, with a mean[median] of 12[9].  We can therefore define
a rough limit on the orbital distance of an inner giant planet that allows
terrestrial planets to form in the Habitable Zone.  For a terrestrial planet
inside the outer edge of the Habitable Zone at 1.5 AU the giant planet's orbit
must be inside roughly 0.5 AU (the most optimistic case puts the giant planet
at 0.68 AU).  We apply this inner giant planet limit to the known sample of
extra-solar giant planets (including planets discovered by the radial
velocity, transit and microlensing techniques \cite{but06,sch02}) in
combination with a previous study of outer giant planets \cite{ray06}.  We
find that 54 out of 158 (34\%) giant planetary systems in our sample permit an
Earth-like planet of at least 0.3 $\mearth$ to form in the Habitable Zone
(Fig. 3).  The fraction of known systems that could be life-bearing may
therefore be significantly higher than previous estimates \cite{ray06}.

The occurrence of Hot Jupiters appears to be a strong function of stellar
metallicity \cite{fis05}.  In addition, the solid component of protoplanetary
disks is assumed to be proportional to metallicity.  Therefore, systems such
as the ones studied here may have very massive solid disks and could have
systematically larger planet masses.  If, for example, such disks are more
likely to form $\sim$ 10 $\mearth$ ``Hot Neptunes'' (e.g., 55 Cnc e
\cite{mcar04}) than $\sim 4 \mearth$ Hot Earths, then our disk is too small by
a factor of a few.  Assuming planet mass scales with disk mass, the typical
mass of a habitable planet in such systems may be several Earth masses.  In
addition, our calculations were for a realistic but fixed giant planet mass
and migration rate.  Less[more] massive giant planets or faster[slower]
migration rates increase[decrease] the survival rate of terrestrial material
exterior to the close-in giant planet \cite{man03}.

Upcoming space missions such as NASA's Kepler and Terrestrial Planet Finder
and ESA's COROT and Darwin will discover and eventually characterize
Earth-like planets around other stars.  We predict that a significant fraction
of systems with a close-in giant planet will be found to have a Hot Earth or
potentially habitable, water-rich planets on stable orbits in the Habitable
Zone.  Suitable targets may be found in the known giant planet systems.

\clearpage
\begin{figure}
\includegraphics[width=15cm]{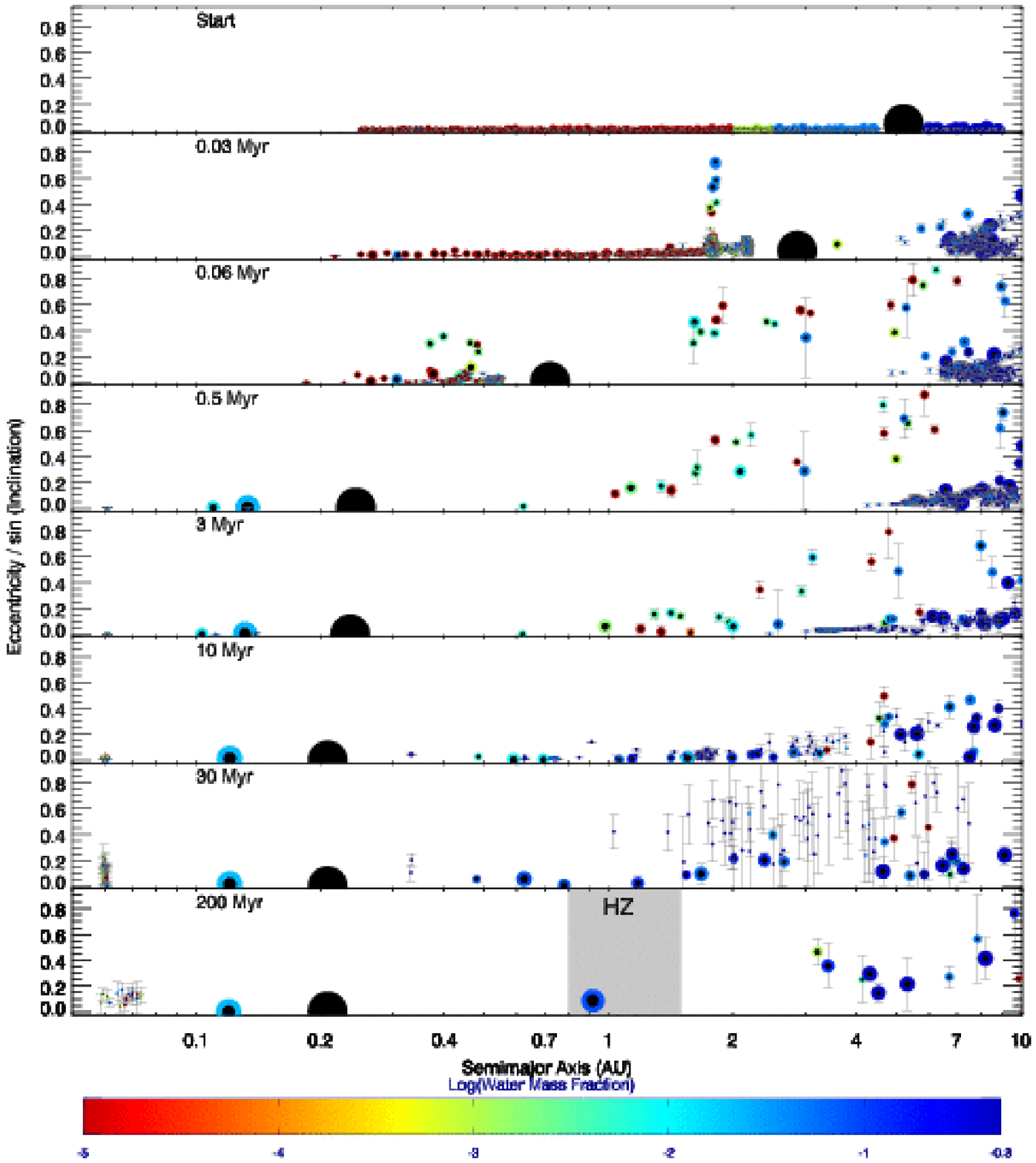}
\caption{{\bf Snapshots in time of the evolution of one simulation}.  Each
panel plots the orbital eccentricity vs. semimajor axis for each surviving
body.  The size of each body is proportional to its physical size (except for
the giant planet, shown in black).  The vertical ``error bars'' represent the
Sine of each body's inclination on the y axis scale.  The color of each dot
corresponds to its water content (as per the color bar), and the dark inner
dot represents the relative size of its iron core.  For scale, the Earth's
water content is roughly $10^{-3}$ \cite{note2}.}
\end{figure}

\begin{figure}
\includegraphics[width=15cm]{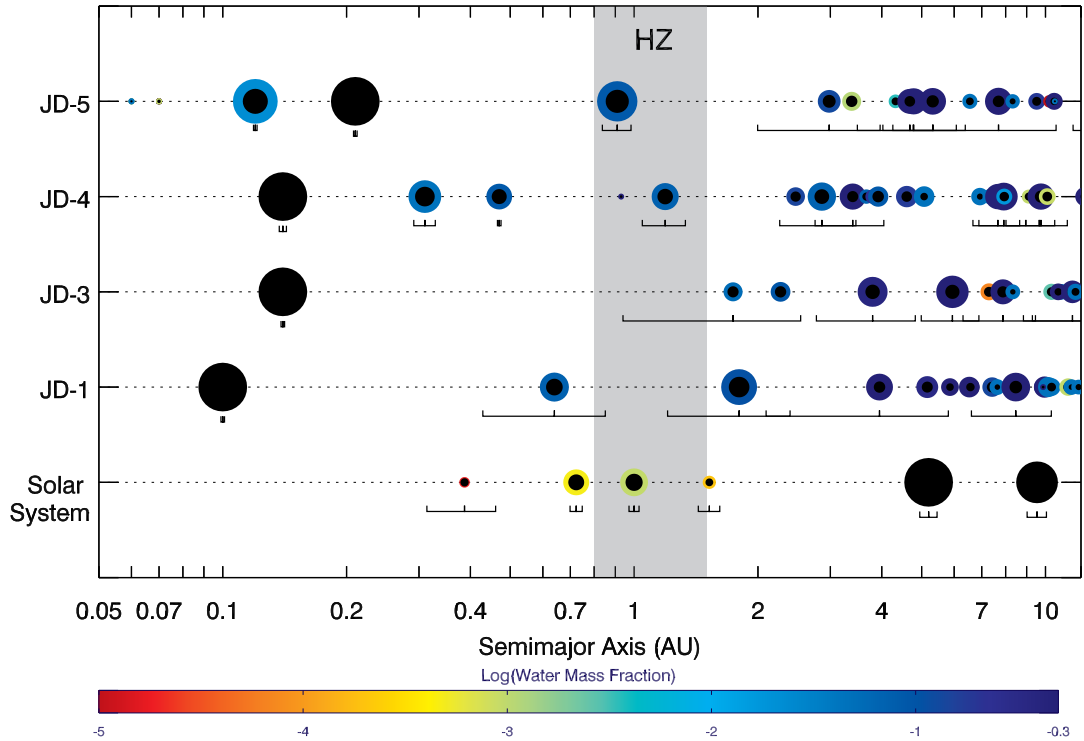}
\caption{{\bf Final configuration of our four simulations, with the Solar
System shown for scale.}  Each simulation is plotted on a horizontal line, and
the size of each body represents its relative physical size (except for the
giant planets, shown in black).  The eccentricity of each body is shown
beneath it, represented by its radial excursion over an orbit.  As in Fig 1,
the color of each body corresponds to its water content, and the inner dark
region to the relative size of its iron core.  The simulation from Fig. 1 is
'JD-5'.  Orbital values are million year averages; Solar System values are 3
million year averages \cite{qui91}.  See Table S1 for details of simulation
outcomes.  Note that some giant planets underwent additional inward migration
after the end of the forced migration, caused by an articial drag force.  This
caused many Hot Earths to be numerically ejected, but had little effect
outside the inner giant planet.  See supporting online material for details.}
\end{figure}

\begin{figure}
\includegraphics[width=15cm]{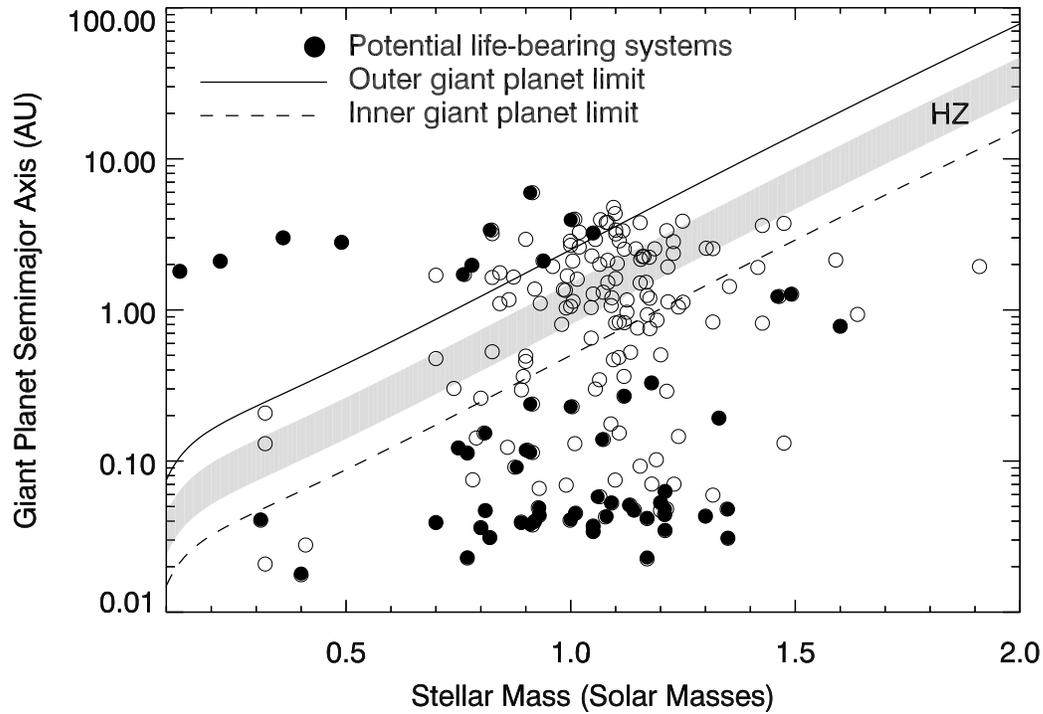}
\caption{{\bf Giant planet orbital parameter space that allows terrestrial
planets to form in the Habitable Zone.}  The solid line indicates the limit
for outer giant planets from \cite{ray06}.  The dashed line is an approximate
limit (0.5 AU with eccentricity less than 0.1 -- the maximum eccentricity
achieved in most simulations -- for a solar-mass star) inside which
low-eccentricity giant planets allow for the formation of habitable planets,
derived from our results and \cite{rql05}.  The Habitable Zone is shaded in
green, and was calculated by assuming the temperature to scale with the
stellar flux (i.e., the square root of the stellar luminosity), using a
stellar mass-luminosity relation fit to data of \cite{hil04}.  Open circles
represent known giant planets that are unlikely to allow habitable terrestrial
planets in the Habitable Zone.  Filled circles represent known planets with
low enough orbital eccentricities to satisfy our criteria for habitable planet
formation, deemed to be potentially life-bearing.}
\end{figure}

\clearpage
\begin{deluxetable}{l|ccccc}
\tablewidth{0pt}
\tablecaption{Properties of Simulated Planets\tablenotemark{1}}
\tabletypesize{\scriptsize}
\tablecolumns{6}
\renewcommand{\arraystretch}{.6}
\tablehead{
\colhead{} &  
\colhead{Hot Earths} & 
\colhead{Normal Terrestrials} & 
\colhead{Hab. Zone planets} &
\colhead{Outer Terrestrials}  &
\colhead{Solar System\tablenotemark{2}}}
\startdata
Mean Number of planets & 0.25\tablenotemark{3} & 2 & 0.5 & 11 & 4\\ 

Mean Planet mass ($\mearth$) & 4.2 & 1.1 & 2.0 & 0.6 & 0.49\\ 

Mean Water mass fraction & $2\times 10^{-2}$ & $8\times 10^{-2}$ & $8\times
10^{-2}$ & $3.5\times 10^{-1}$ & $4\times 10^{-4}$\\

Mean Iron mass fraction & 0.25 & 0.28 & 0.27 & 0.14 & 0.32\tablenotemark{4}\\

Mean Orbital eccentricity & 0.01 & 0.23 & 0.10 & 0.23 & 0.08\\ 

Mean Orbital inclination (deg) & 0.7 & 11 & 7 & 13 & 3.0\\

\enddata
\tablenotetext{1}{We include results from four simulations, described in
detail in the supporting online material.}
\tablenotetext{2}{Solar System physical properties from \cite{lod98} and
orbital properties from \cite{qui91}.}
\tablenotetext{3}{Every simulation with gas drag formed 1-3 Hot Earths during
  giant planet migration.  However, in many cases an artificial drag force
  caused continued inward migration of the giant planet.  In these cases, the
  integrator usually introduced an error causing the eventual ejection of Hot
  Earths when they entered $\sim$0.05 AU.  We therefore consider 0.5 a lower
  bound on the frequency of Hot Earths in these systems.  See also
  \cite{fog05}.}
\tablenotetext{4}{Solar System iron mass fraction are calculated without Mercury,
  because of its anomalously high iron content.}
\end{deluxetable}


\begin{thebibliography}{99}
\bibitem{note1} An alternative model
for the presence of Hot Jupiters contends that they were dynamically scattered
to small orbital distances, where tidal effects circularized their orbits
\cite{ras96}.  If that process is in fact dominant over the type 2
migration model considered here, then these models need to be re-considered.
During the very high-eccentricity phase, it is likely that the giant planet
could remove most of the terrestrial mass in the system, leaving little
remaining to form planets.
\bibitem{note2} The Earth's total water content is between roughly 2 and 10
``oceans'' (see \cite{lec98}), where 1 ocean is the amount of water on our
planet's surface (1.5$\times10^{24}$ grams); Earth is about 0.1\% water by
mass.  Note also that although we did not account for water loss during
impacts, water contents of these simulated planets are more than an order of
magnitude higher than in previous simulations \cite{rql04}.  So although the
water contents listed in Table 1 are upper limits, the very high values are
real.
\end{thebibliography}
\end{document}